# Theoretical study on the electronic properties and multiorbital models of La$_3$Ni$_2$O$_7$ thin films on SrLaAlO$_4$ (001)

Guanlin Li[1†], Cui-Qun Chen[2†], Haoliang Shi[1], Zhengtao Liu[1], Hao Ma[1], Fubo Tian[1,*], Dao-Xin Yao[2,*], Defang Duan[1,*]

[1] Key Laboratory of Material Simulation Methods & Software of Ministry of Education, State Key Laboratory of High Pressure and Superhard Materials, College of Physics, Jilin University, Changchun 130012, China

[2] Center for Neutron Science and Technology, Guangdong Provincial Key Laboratory of Magnetoelectric Physics and Devices, State Key Laboratory of Optoelectronic Materials and Technologies, School of Physics, Sun Yat-Sen University, Guangzhou, 510275, China.

*Corresponding authors: tianfb@jlu.edu.cn (F.T.); yaodaox@mail.sysu.edu.cn (D-X.Y.); duandf@jlu.edu.cn (D.D)

†These authors contributed equally to this work.

## Abstract

The realization of high-temperature superconductivity in La$_3$Ni$_2$O$_7$ thin films at ambient pressure has triggered a fundamental inquiry into the microscopic origin of the metallic ground state—specifically, whether the requisite charge carrier density and orbital polarization are induced purely by lattice geometry (strain), or necessitate an intrinsic charge reconstruction driven by the interface. Here, we present a systematic density functional theory (DFT+$U$) study of La$_3$Ni$_2$O$_7$/SrLaAlO$_4$ (001) heterostructures, utilizing a fully self-consistent model that explicitly incorporates substrate-induced compressive strain and interfacial reconstruction. We identify the intrinsic hole doping induced by interfacial Sr interdiffusion as the decisive factor in stabilizing the experimentally observed metallic ground state. Crucially, our 1-unit-cell (UC) heterostructure model naturally reproduces the detailed Fermi surface topology observed in angle-resolved photoemission spectroscopy (ARPES), most notably the critical Ni-$d_{z^2}$ derived $\gamma$ hole pocket at the Fermi level ($E_F$). Layer-resolved analysis further reveals that this conductive $\gamma$ band originates exclusively from the interface-proximal bilayer, suggesting that macroscopic transport is dominated by the interfacial region. Based on this correspondence, we construct a minimal double-stacked two-orbital tight-binding model. Comparative analysis with the bulk phase suggests that the reduction in superconducting transition temperature ($T_c$) in thin films is attributable to the combined effects of the significantly reduced electronic density of states (DOS) at the Fermi level and the suppression of vertical superexchange coupling ($J_\perp^Z$), driven by strain-induced out-of-plane lattice expansion. Our findings establish that interface engineering goes beyond mere strain imposition, acting as a decisive factor in modulating nickelate orbital physics, thus providing a rigorous framework for understanding ambient-pressure superconductivity in thin-film devices.

# 1    Introduction

Nearly four decades after the discovery of cuprate superconductors [1-5], a breakthrough emerged in the nickelate family—element adjacent to copper in the periodic table—with the observation of superconductivity in square-planar infinite-layer thin films [6]. This phenomenon was first demonstrated in $Nd_{0.8}Sr_{0.2}NiO_2$ films epitaxially grown on atomically flat $SrTiO_3$ (001) substrates. These materials exhibit striking similarities to copper oxides [7-10] in both their crystal and electronic structure: their crystal lattices follow the transition metal-oxygen plane structural pattern, while their electronic structure is characterized by a nominal $3d^9$ valence configuration. Consequently, the prevailing consensus suggests that the unconventional superconductivity of infinite-layer nickel oxides primarily originates from the Ni-$d_{x^2-y^2}$ band [11-17]. Subsequent studies revealed that high-temperature superconductivity with a $T_c$ of 80 K was discovered in the double-layer nickelate $La_3Ni_2O_7$ (LNO) under high pressure (approximately 14 GPa) [18-21]. This structure is a member of the Ruddlesden-Popper (RP) family, described by the general formula $A_{n+1}B_nO_{3n+1}$ ($n=2$). This breakthrough surpassed the limitations of liquid-nitrogen temperatures and marked a significant milestone in condensed-matter physics. In contrast to copper oxides and infinite-layer nickel oxides, the nickel ions in LNO possess a $3d^{7.5}$ electronic configuration, exhibiting enhanced complexity stemming from contributions of the Ni-$d_{z^2}$ orbital. Research indicates that the pressure-driven emergence of superconductivity coincides with a structural transition to a bilayer orthorhombic (*Fmmm*) or tetragonal (*I4/mmm*) phase and a Lifshitz phase transition involving the $\gamma$ band derived from Ni-$d_{z^2}$ bonding orbitals, where interlayer coupling has been identified as a potential critical factor in the process of superconducting pairing [22-44]. However, unraveling the precise pairing mechanism remains elusive, stemming largely from the experimental challenges inherent to characterizing complex quantum states within the constraints of high-pressure environments.

Epitaxial thin-film growth offers a promising route to stabilize these high-pressure superconducting states through heterostructural engineering. By utilizing substrate-imposed biaxial strain and interfacial coupling, it is possible to not only tune the lattice parameters but also modulate oxygen octahedral rotations within the perovskite framework [45-49]. A recent experiment demonstrated that, under tensile strain, thin films are more prone to form perovskite structures ($n = \infty$), whereas compressive strain stabilizes the LNO phase ($n = 2$) [50]. Building on this, two independent research groups nearly simultaneously discovered that $SrLaAlO_4$ (SLAO) (001) substrates provide a distinct platform for stabilizing the superconductivity of LNO thin films at ambient pressure, achieving a $T_c$ exceeding the empirical McMillan limit (approximately 42 K) [51,52]. This approach circumvents the requirement for external mechanical pressure, a significant impediment in practical applications, thereby establishing SLAO as a transformative material platform [53-55]. Mechanistically, the SLAO substrate imposes a -2.0% biaxial compressive strain, mimicking the structural environment of bulk LNO under ~14 GPa hydrostatic pressure. The resulting straightening of Ni-O-Ni bonds (168°→180°) and the suppression of octahedral tilting enhance the electronic bandwidth and reduce the crystal-field splitting, thereby promoting the orbital degeneracy between $d_{z^2}$ and $d_{x^2-y^2}$ states. Crucially, this ambient-pressure stabilization grants direct access to the electronic structure via ARPES, allowing for experimental verification of the relevant orbital physics. However, the experimental picture regarding the critical $\gamma$ band remains contentious: while initial ARPES studies on compressively strained $La_{2.85}Pr_{0.15}Ni_2O_7$ and $La_{2.31}Pr_{0.24}Sm_{0.45}Ni_2O_7$ thin films [56,57] corroborated the presence of the density functional theory (DFT) predicted $\gamma$ band Fermi pocket, recent ARPES measurements on superconducting $La_2PrNi_2O_7$ and $La_{2.79}Sr_{0.21}Ni_2O_7$ thin films reveal Fermi surface topologies incompatible with the existence of the $\gamma$ band [58,59]. This complex and seemingly contradictory experimental landscape underscores the sensitivity of the electronic ground state to the specific thin-film environment.

In parallel with these experimental explorations, extensive theoretical studies have been performed to understand the electronic structure and pairing symmetry of LNO thin films [60-71]. However, the prevailing theoretical frameworks largely rely on simplified approximations, ranging from biaxially strained bulk models to free-standing slab calculations. While these approaches successfully capture the broad similarity to the high-pressure phase and the presence of the $\alpha$ and $\beta$ electron pockets, they often encounter difficulties in self-consistently determining the critical hole-doping level and the exact energy alignment of the Ni-$d_{z^2}$ derived $\gamma$ band. Such treatments inherently lack a self-consistent description of the charge transfer effects specific to the film-substrate interface and fail to account for the interfacial reconstruction. Crucially, recent scanning transmission electron microscopy (STEM) and energy-dispersive X-ray spectroscopy (EDS) measurements have explicitly resolved interfacial reconstruction and Sr interdiffusion across the LNO/SLAO interface [55-57]. The central unresolved scientific question, therefore, is whether the geometric strain effect alone is sufficient to replicate the electronic ground state characteristics of the superconducting phase, or if the intrinsic charge reconstruction driven by interfacial chemistry plays a determinative role. Addressing this requires a rigorous heterostructure framework that transcends free-standing or strained-bulk approximations to elucidate the realistic interplay between lattice strain and interfacial doping.

In this paper, we present a systematic first-principles investigation of the electronic and structural evolution of RP nickelates from 0.5 to 5 UC LNO on SLAO. Distinct from previous studies relying on free-standing slab approximations, our approach employs a fully self-consistent heterostructure model that explicitly incorporates substrate-induced strain and interfacial reconstruction. We first demonstrate that Sr interdiffusion acts as an intrinsic chemical doping mechanism, providing a natural explanation for the hole-doped metallic character observed in these films. By comparing films of varying thicknesses, we show that the calculated fermiology of the full-UC configurations (except 0.5 UC)—most notably the 1 UC limit—achieves excellent quantitative agreement with ARPES measurements, successfully capturing the complete set of $\alpha$, $\beta$, and $\gamma$ pockets. Layer-resolved analysis further reveals that the conductive $\gamma$ band near $E_F$ originates exclusively from the interface-proximal bilayer (Stack 1), highlighting the dominant role of the interface in macroscopic transport. Finally, motivated by this structural and electronic correspondence, we derive a minimal double-stacked two-orbital tight-binding model from the 1 UC system to capture the essential low-energy physics, offering a unified theoretical framework for understanding ambient-pressure superconductivity in nickelate thin films.

## 2   Computational details

First-principles calculations were performed within the DFT framework, implemented in the Vienna ab initio simulation package [72]. The exchange-correlation terms were described using the generalized gradient approximation (GGA) in the Perdew-Burke-Ernzerhof (PBE) form [73]. Projector-augmented wave (PAW) potentials were employed [74,75], with the valence shells of Ni, La, Sr, and O treated as $3s^23p^63d^84s^2$, $5s^25p^65d^16s^2$, $4s^24p^65s^2$, and $2s^22p^4$, respectively. The plane-wave cutoff energy was set to 600 eV. The Brillouin zone was sampled using $\Gamma$-centered Monkhorst-Pack $k$-point meshes with densities of $2\pi \times 0.03$ Å$^{-1}$ for structural relaxation, and $2\pi \times 0.02$ Å$^{-1}$ for self-consistent and DOS calculations. In thin-film form, LNO stabilizes in a tetragonal structure analogous to its high-pressure bulk phase, characterized by near-linear Ni-O-Ni bond angles (~180°) [51]. Accordingly, thin-film models with thicknesses ranging from 0.5 to 5 UC were constructed by truncating the experimentally determined bulk $I4/mmm$ phase along the (001) direction. To simulate epitaxial growth on SLAO substrates, the in-plane lattice parameters were constrained to $a=b=3.754$ Å [52,60], inducing a compressive epitaxial strain of $\varepsilon=a_{SLAO}/a_{LNO}-1\approx-2.0\%$, where $a_{LNO}=3.837$ Å is the ambient-pressure lattice constant of bulk LNO [41]. The out-of-plane lattice constant ($c$) and internal atomic coordinates were fully relaxed until the Hellmann-Feynman forces on each atom converged to less than 0.005 eV/Å. The energy convergence criterion for the electronic self-consistency loop was set to $10^{-8}$ eV. A vacuum region of ~20 Å was included along the z-direction to eliminate spurious interactions between periodic images. To account for the significant Sr interdiffusion observed experimentally across approximately 1 UC at the interface, the cation stoichiometry was modeled using the virtual crystal approximation (VCA) [76,77]. Electronic correlation effects were treated using the DFT+$U$ [78,79] with an on-site Coulomb repulsion of $U = 3.5$ eV applied to the Ni $3d$ orbitals [80]. To construct low-energy effective models, maximally localized Wannier functions (MLWF) are obtained using the Wannier90 code [81-83].

## 3      Results and discussions

To systematically investigate the dimensional evolution of the heterostructures, we modeled LNO thin films with thicknesses of 0.5, 1, 2, 3, 4, and 5 UC. For clarity in the subsequent discussion, we explicitly categorize the 0.5 UC case as the "half-UC" configuration, while the integer-thickness films (1-5 UC) are collectively designated as "full-UC" configurations. Figure 1(a) shows the optimized geometric structure of a representative 1 UC LNO film on a 2 UC SLAO (001) substrate, while the configurations for the 0.5 UC case and thicker films (2-5 UC) are provided in the Supporting Information (Figure S1). The LNO/SLAO(001) heterostructure model consists of three principal components: a LNO film layer with a variable thickness cleaved from the high-pressure $I4/mmm$ phase, an interface reconstruction layer consistent with STEM observations [56], and a 2 UC SLAO substrate layer. To account for the solid-solution nature of the cations, the chemical disorder at the perovskite $A$-sites (corresponding to La sites in the film and mixed Sr/La sites in the substrate) is treated using VCA. The validity of this VCA approach for the substrate layer (Sr: La = 1:1) was explicitly confirmed via supercell calculations (Figure S2). To capture the realistic charge redistribution, the $A$-site cations across the substrate, the reconstruction layer, and specifically the interface-proximal 1 UC of the film were modeled using VCA with three distinct Sr: La mixing ratios. These ratios were rigorously determined by enforcing charge neutrality across the structural transition from the $A_2BO_4$-type structure (SrLaAlO$_4$) to the $A_3B_2O_7$-type topology, which takes place within the interface reconstruction layer. This methodology effectively models the natural diffusion of Sr while respecting the physical observation that such interdiffusion is confined to the first unit cell of the LNO film [52,56]. Consequently, for films thicker than 1 UC, the upper layers are treated as stoichiometric LNO, ensuring that the derived mixing ratios represent a rigorous chemical boundary condition rather than arbitrary doping parameters.

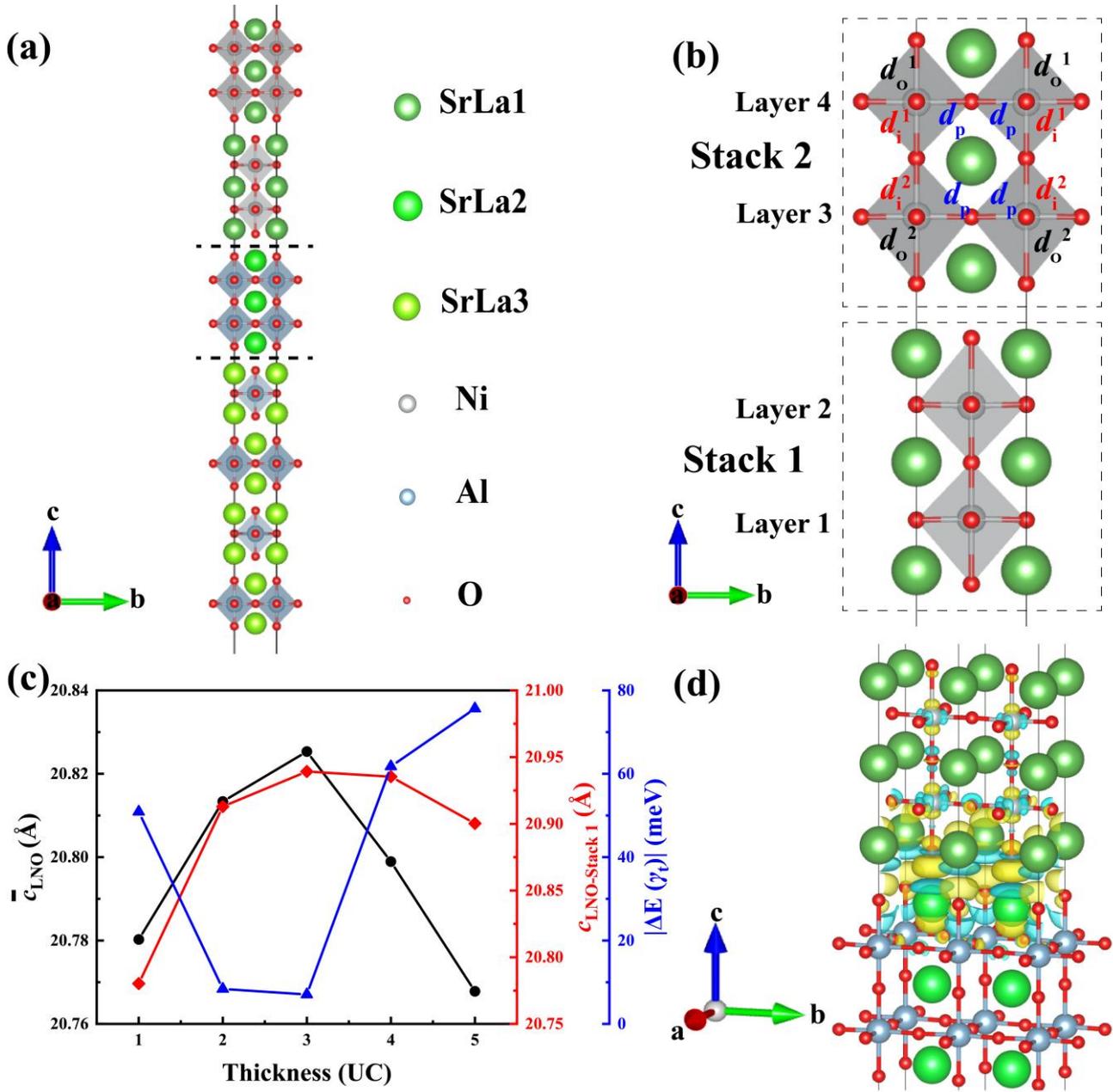

**Figure 1** (a) Optimized geometry of 1 UC LNO on SLAO substrate. (b) The two bilayers are denoted as Stack 1 and Stack 2, with Stack 1 at the substrate interface and Stack 2 farther from the interface, adjacent to the vacuum layer. Key Ni-O bond lengths are indicated: outer-apical ($d_o^1$, $d_o^2$; black arrows), inner-apical ($d_i^1$, $d_i^2$; red arrows), and in-plane ($d_p$; blue arrows). Three distinct Sr: La doping ratios are schematically represented by different shades of green spheres. (c) Thickness-dependent trends of the average out-of-plane lattice parameter $\bar{c}_{LNO}$, the height of the interface-proximal Stack 1 ($c_{LNO\text{-}Stack\,1}$), and the absolute energy separation $|\Delta E\,(\gamma_t)|$ between the top of the $\gamma_t$ bonding band and $E_F$. (d) Charge density differences $\Delta\rho$ of heterostructure, defined as $\Delta\rho$ ($\Delta\rho =\rho_{system}- \rho_{LNO}- \rho_{substrate}$), plotted with an isosurface level of ± 0.015 e/Å$^3$. Yellow and blue regions correspond to electron accumulation and depletion, respectively.

Experimental investigations frequently employ isovalent lanthanide substitution, a strategy originally established in studies of the high-pressure compound $La_2PrNi_2O_7$ [21,84] to introduce chemical pressure and enhance crystalline quality. While various experimental compositions exist, including stoichiometric $La_2PrNi_2O_7$ films [53], our theoretical analysis specifically benchmarks against the $La_{2.85}Pr_{0.15}Ni_2O_7$ films [52] used in recent breakthroughs. In this specific case, only 5% of the *A*-sites are substituted with trace amounts of Pr. Our calculations confirm that such minimal Pr doping induces negligible perturbations to the electronic structure near $E_F$ (Figure S3). This finding validates our approximation of treating the upper film layers as stoichiometric LNO, allowing the model to focus exclusively on the La/Sr framework.

Table I The Ni-O bond lengths (*d*) in $NiO_6$ octahedra of LNO/SLAO heterostructure. The unit of *d* is in Å.

| Stack | d | 0.5 UC | 1 UC | 2 UC | 3 UC | 4 UC | 5 UC |
|---|---|---|---|---|---|---|---|
| 1 | $d_o^1$ | 2.077 | 2.330 | 2.316 | 2.322 | 2.321 | 2.316 |
|  | $d_i^1$ | 1.955 | 1.962 | 1.959 | 1.961 | 1.961 | 1.959 |
|  | $d_i^2$ | 1.969 | 1.939 | 1.940 | 1.941 | 1.940 | 1.939 |
|  | $d_o^2$ | 2.167 | 2.245 | 2.245 | 2.248 | 2.246 | 2.239 |
|  | $d_p$ | 1.877 | 1.877 | 1.877 | 1.877 | 1.877 | 1.877 |
| 2*n* | $d_o^1$ | - | 2.131 | 2.135 | 2.131 | 2.132 | 2.132 |
|  | $d_i^1$ | - | 1.936 | 1.924 | 1.923 | 1.923 | 1.923 |
|  | $d_i^2$ | - | 1.999 | 2.029 | 2.032 | 2.030 | 2.030 |
|  | $d_o^2$ | - | 2.316 | 2.401 | 2.407 | 2.401 | 2.400 |
|  | $d_p$ | - | 1.877 | 1.877 | 1.877 | 1.877 | 1.877 |

As shown in Figure 1(b), the representative 1 UC film consists of two vertically stacked $NiO_2$ bilayers, designated as Stack 1 (interface-proximal) and Stack 2 (surface-proximal). This nomenclature generalizes to any generic *n* UC films, where the bilayer adjacent to the vacuum is consistently denoted as Stack 2*n*. Key structural parameters, specifically the outer-apical ($d_o^1$, $d_o^2$), inner-apical ($d_i^1$, $d_i^2$), and in-plane ($d_p$) Ni-O bond lengths, are defined in the figure and summarized by stack index in Table I. For clarity, only the boundary bilayers—Stack 1 and Stack 2*n*—are explicitly labeled. Across all thicknesses, the hierarchy of bond lengths remains robust: outer-apical Ni-O bonds ($d_o^1$, $d_o^2$) are systematically longer than inner-apical Ni-O bonds ($d_i^1$, $d_i^2$), a feature consistent with the structural characteristics of the high-pressure bulk LNO phase [29]. However, substrate interactions break the local symmetry of the interlayer Ni-O bonds. In the interface-proximal Stack 1 of all full-UC configurations, this results in the inequalities $d_o^1 > d_o^2$ and $d_i^1 > d_i^2$. This asymmetry attenuates with increasing distance from the interface and reverses in the vacuum-adjacent Stack 2*n*, attributed to the absence of vertical confinement at the free surface. In contrast, the half-UC configuration exhibits fundamentally distinct bonding characteristics arising from its unique symmetry constraints compared to the full-UC limit. We anticipate that these structural deviations will significantly influence the electronic structure, as discussed below. By incorporating both substrate-induced strain and Sr interdiffusion, our heterostructure model provides a more rigorous structural foundation than simplified free-standing approximations, thereby ensuring an accurate baseline for the subsequent analysis of electronic and superconducting properties.

The total film thickness, $c_{LNO}$, is defined as the vertical distance from the substrate interface to the topmost LNO atomic layer. To quantify the vertical structural evolution, we define the average out-of-plane lattice parameter, $\bar{c}_{LNO}$, by dividing $c_{LNO}$ by the number of unit cells ($n$). As plotted in Figure 1(c), the evolution of $\bar{c}_{LNO}$ across the full-UC configurations exhibits two distinguishing features. First, the film undergoes a pronounced out-of-plane elongation driven by the compressive epitaxial strain (Poisson effect). For the 1 UC case, $\bar{c}_{LNO}$ reaches 20.780 Å, notably exceeding the $c$-axis lattice constant of the high-pressure bulk phase at 21.6 GPa (19.766 Å) [41]. This strain-induced expansion increases the interlayer spacing, which likely weakens the interlayer superexchange coupling. This mechanism offers a plausible explanation for the suppressed $T_c$ observed in compressively strained films compared to their pressurized bulk counterparts. Second, $\bar{c}_{LNO}$ displays a non-monotonic thickness dependence: it increases continuously up to 3 UC, beyond which it begins to decline. This structural trend is inversely correlated with the electronic parameter $|\Delta E (\gamma_t)|$, defined as the absolute energy separation between the band top of the highest-lying Ni-$d_{z^2}$ derived bonding band ($\gamma_t$) and $E_F$. The strong correlation indicates a tight coupling between the vertical lattice dimension and the proximity of the Ni-$d_{z^2}$ bonding states to the Fermi surface. A parallel structural evolution is observed in the interface-proximal Stack 1 ($c_{LNO\text{-}Stack\ 1}$). Given the established role of the $d_{z^2}$ bonding band in the superconductivity of bilayer nickelates, these findings suggest that the superconductivity of these epitaxial films is critically sensitive to the local electronic and structural properties of Stack 1.

We further elucidated the electronic reconstruction at the atomic limit by calculating the planar-averaged charge density difference ($\Delta\rho$) along the z-direction, as visualized in Figures 1(d) and S4. The profile reveals pronounced charge density oscillations across the interface, characterized by significant electron depletion (negative $\Delta\rho$) at the Ni-O metallic planes and concurrent electron accumulation (positive $\Delta\rho$) at the adjacent SrLa1-O spacer layers. This specific local redistribution signifies that electrons are transferred from the Ni-O orbitals to the dopant-hosting $A$-site layers, consistent with the mechanism of hole doping induced by divalent Sr substitution. To quantify the net charge transfer, we performed Bader charge analysis on the heterostructure. Summing the Bader charges over the entire LNO film region reveals a negligible net charge transfer from the substrate ($|\Delta Q| < -0.02$ e), corroborating the absence of long-range charge injection. Collectively, these features confirm that the metallic ground state of the LNO film is not driven by extrinsic interfacial charge transfer, but rather stems principally from intrinsic chemical doping mediated by the Sr interdiffusion.

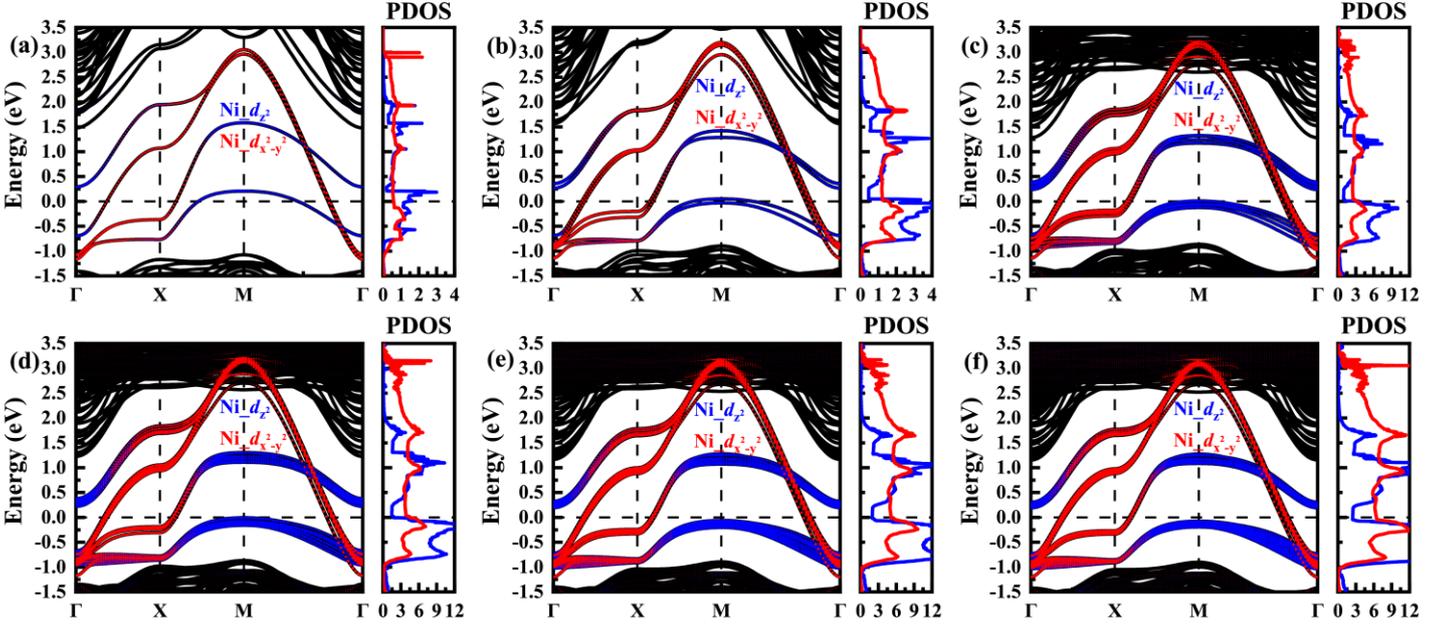

**Figure 2** Orbital-projected band structures and DOS for LNO films at thicknesses of (a) 0.5, (b) 1, (c) 2, (d) 3, (e) 4, and (f) 5 UC. ($U$ = 3.5 eV) Energy-resolved orbital projections are shown along high-symmetry directions: blue and red colors correspond to Ni-$d_{z^2}$ and Ni-$d_{x^2-y^2}$ orbital contributions, respectively.

We investigated the electronic ground states by calculating the band structures and projected density of states (PDOS) of LNO thin films using DFT+$U$ with an effective Hubbard parameter of $U$ = 3.5 eV. All systems exhibit metallic behavior, with the electronic states near $E_F$ being dominated by Ni-$d_{x^2-y^2}$ and Ni-$d_{z^2}$ orbitals, consistent with the electronic structure of bulk LNO [25] (Figure 2). Notably, in the full-UC films, the Ni-$d_{z^2}$ derived $\gamma$ bonding band splits along the Γ-M and X-M directions due to the reduced structural symmetry imposed by the interface, whereas the single-bilayer nature of the half-UC structure results in a single, unsplit $\gamma$ band [Figure 2(a)]. For the split bands, we approximate the band center, $\gamma_c$, by averaging the eigenenergies of the highest and lowest subbands. To quantify the band evolution, we define the parameter |ΔE ($\gamma_c$)| as the absolute energy separation between $\gamma_c$ and $E_F$. As the film thickness increases across the full-UC configurations, the entire $\gamma$ bonding band shifts to lower energies relative to $E_F$, a trend captured by the monotonic increase in |ΔE ($\gamma_c$)| (Figure S5). Crucially, the Ni-$d_{z^2}$ derived $\gamma$ bands cross $E_F$ only in thinner films (0.5-2 UC). In thicker films, these bands lie slightly below $E_F$; for instance, in the 3 UC model, the $\gamma_t$ band is positioned 7 meV below $E_F$. This energetic evolution directly dictates the DOS characteristics: the sinking of the prominent $\gamma$ band peak suppresses the PDOS at $E_F$, while the submergence of high-energy band tails increases the electron occupancy, as evidenced by the rising integrated projected density of states (IPDOS) shown in Figure S5. This behavior stems fundamentally from the hole dilution effect imposed by the boundary conditions, which maintains a fixed interfacial Sr reservoir. As the number of Ni layers increases, the fixed Sr content results in a progressive dilution of the effective hole doping per Ni site. This reduction in hole concentration necessitates an increase in electron filling, which is electronically realized by the downward shift of the $\gamma$ bands relative to $E_F$. It might be hypothesized that in experimental setups, substrates act as semi-infinite reservoirs where annealing could induce enhanced Sr diffusion, potentially pinning the $\gamma$ band at the Fermi surface even in thicker films. We tested this scenario by incorporating enhanced Sr diffusion from the interface reconstruction layer to the LNO film layer into the heterostructure model under the constraint of fixed total Sr content (Figure S6). Counterintuitively, our results indicate that as Sr doping concentration increases, the energy of the $\gamma$ band originating from Stack 1—originally closest to $E_F$—gradually decreases. Excessive Sr doping significantly disrupts LNO structural integrity, thereby validating the physical principles underlying our valence balance strategy.

Despite the general trend of decreasing PDOS with thickness, the 1 UC film exhibits the maximum spectral weight among the series (0.652 states/eV/Ni), contrasting with thicker films. This peak intensity arises from the extreme two-dimensional quantum confinement inherent to the single-unit-cell limit, where the absence of inter-unit-cell vertical coupling suppresses dispersion along the $k_z$ direction, flattening the $\gamma$ band into a sharp DOS peak. In contrast, for thicker films, the onset of inter-unit-cell coupling induces band splitting and broadening, which effectively dilutes the spectral weight over a wider energy range. However, a critical comparison reveals that even this maximum film PDOS remains significantly lower than that of the superconducting high-pressure bulk phases: 1.333 states/eV/Ni for *Fmmm* phase (21.6 GPa) and 1.047 states/eV/Ni for *I*4/*mmm* phase (60.3 GPa) (Figure S7). Furthermore, layer-resolved PDOS of Ni-$d_{z^2}$ shown in Figure S8 further reveals that consistently across the full-UC series, the $\gamma$ band closest to $E_F$ originates exclusively from Ni atoms in Stack 1. This implies that the macroscopic conductivity, and potentially the superconductivity, is dominated by the interfacial layer, aligning with experimental observations linking strong superconductivity to different interfaces [51-53].

Previous studies suggest that a lower Hubbard $U$ ($\leq 2$ eV) better captures the experimentally observed spin-density-wave (SDW) order, with prior thin-film studies employing $U = 2.0$ eV [62,85]. For comparison, we performed parallel calculations at the $U = 2.0$ eV case and found that the $\gamma$ band intersects $E_F$ across all film thicknesses (Figure S9). We emphasize that regardless of whether the $\gamma$ band strictly crosses $E_F$, its energetic proximity facilitates Ni-$d_{z^2}$ charge fluctuations, thereby promoting interlayer antiferromagnetic coupling and providing a viable pathway to superconductivity. These findings highlight the critical role of interfacial engineering in modulating electronic states near $E_F$, offering a unified theoretical framework for understanding high-temperature superconductivity in epitaxial nickelate films.

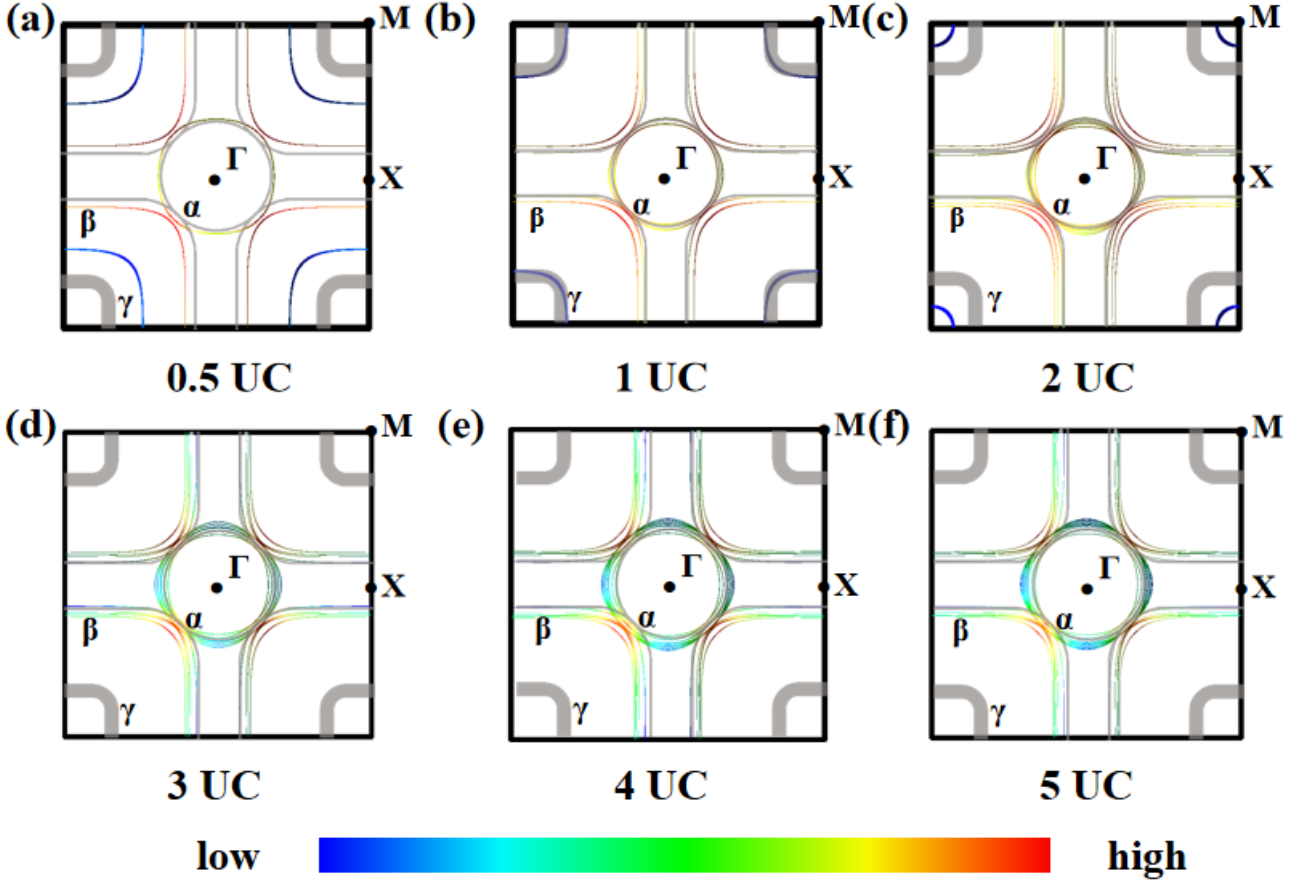

**Figure 3** Evolution of the Fermi surface geometry in LNO/SLAO(001) thin films. Panels (a)-(f) display the calculated Fermi surfaces for film thicknesses of 0.5, 1, 2, 3, 4, and 5 UC, respectively, obtained using DFT+$U$ ($U$ = 3.5 eV). The theoretical results are visualized as color-mapped contours, where the hue represents the local magnitude of $v_F$; cool colors (blue/cyan) indicate low $v_F$ (signifying flat-band character with heavy effective mass), while warm colors (yellow/red) indicate high $v_F$ (signifying highly dispersive bands). The distinct Fermi sheets are labeled as $\alpha$ (central electron pocket), $\beta$ (outer hole-like sheets), and $\gamma$ (corner hole pockets). For comparison, experimentally measured Fermi surface contours derived from ARPES data are superimposed as thick gray lines.

The Fermi surface of bulk LNO is characterized by two dominant features: the $\alpha$ sheet, derived primarily from the Ni-3$d_{x^2-y^2}$ orbital character, and the $\beta$ sheet, which exhibits mixed character with partial contributions from Ni-3$d_{z^2}$ states [22-24,80]. Under hydrostatic pressure, the system undergoes a significant reconstruction of its fermiology, wherein the metallization of Ni-3$d_{z^2}$ states leads to the emergence of a $\gamma$ hole pocket [18,22-28]. Figure 3 illustrates the evolution of the Fermi surface geometry in strained LNO/SLAO(001) films, revealing a striking resemblance to the bulk electronic structure. In these plots, the color gradient represents the Fermi velocity ($v_F$) magnitude, offering direct insight into the effective mass distribution. Specifically, the Fermi surface topology clearly resolves the distinct $\alpha$ and $\beta$ sheets. The $\alpha$ sheet exhibits a quasi-isotropic circular geometry with high Fermi velocities (warm colors), characteristic of the itinerant Ni-$d_{x^2-y^2}$ states. In contrast, the $\beta$ sheet displays pronounced warping along the $\Gamma$-M direction, reflecting strong hybridization with Ni-$d_{z^2}$ orbitals. This hybridization culminates in the emergence of the $\gamma$ hole pockets at the M points (in 0.5-2 UC), which are characterized by significantly suppressed Fermi velocities (cool colors). This suppression signifies a flat-band character with heavy effective mass, directly corresponding to regions of high local density of states that are critical for fostering electronic instabilities. At $U$ = 3.5 eV, the $\gamma$ hole pocket is preserved only in 0.5, 1, and 2 UC configurations. While the 0.5 UC film qualitatively captures the presence of the $\gamma$ pocket at $E_F$, the detailed geometric contours of the Fermi sheets—spanning the $\alpha$, $\beta$, and $\gamma$ sectors—exhibit noticeable deviations from ARPES observations. In contrast, full-UC models yield $\alpha$ and $\beta$ pocket trajectories that are highly consistent with experimental data. Notably, the 1 UC film reproduces the $\gamma$ pocket features with strong fidelity to ARPES data, despite minor deviations from the experimentally observed rectangular geometry. This agreement is particularly significant given that 1 UC samples currently yield the highest-quality spectral data in experiments. When the Hubbard $U$ is reduced to 2.0 eV, the Ni-$d_{z^2}$ bonding band shifts upward toward $E_F$, leading to reappearance of the $\gamma$ hole pocket near the M point even in thicker films (Figure S10). Conversely, the geometry of the half-UC films remains largely insensitive to this parameter. To our knowledge, this work presents the first fully self-consistent theoretical study to quantitatively reproduce the experimental ARPES Fermi surface geometry of LNO thin films.

Motivated by the qualitative similarity in electronic structure observed across atomic-layer films of varying thicknesses, we construct a minimal tight-binding model based on the representative 1 UC system to capture the essential band-structure characteristics of the epitaxial series. Utilizing Wannier downfolding of the DFT calculated electronic structure, we establish a double-stacked, two-orbital Hamiltonian incorporating both Ni-$d_{x^2-y^2}$ and Ni-$d_{z^2}$ orbitals written as follows:

$$H = \sum_{i\alpha\sigma} \epsilon_\alpha \hat{c}^\dagger_{i\alpha\sigma} \hat{c}_{i\alpha\sigma} + \sum_{ij,\alpha\beta,\sigma} t^{\alpha\beta}_{ij} (\hat{c}^\dagger_{i\alpha\sigma} \hat{c}_{j\beta\sigma} + h.c.). \qquad (1)$$

Here, $i/j$, $\alpha/\beta$, and $\sigma$ indicate the indices of site, orbital ($d_{x^2-y^2}$ and $d_{z^2}$), and spin, respectively. $\varepsilon_\alpha$ represents the on-site energy of orbital $\alpha$. The band structure from the effective model exhibits good agreement with the underlying DFT results, as shown in Figure 4(a). We consider hoppings up to the third-nearest neighbor, and the parameters are summarized in Table II.

Table II Tight-binding parameters for the 1 UC LNO film model. The superscript notation follows: the on-site energies for the Ni-$d_{x^2-y^2}$ and Ni-$d_{z^2}$ orbitals in layers are denoted by $t_{[00]}^x$ and $t_{[00]}^z$, respectively. x indicates hopping between $d_{x^2-y^2}$ orbitals, z indicates hopping between $d_{z^2}$ orbitals, and xz indicates hopping between $d_{x^2-y^2}$ and $d_{z^2}$ orbitals. Values in parentheses correspond to parameters derived from Wannier downfolding of DFT calculations for the high-pressure bulk phase [25]. All values are eV.

| Layer | i | j | $t_{[i,j]}^x$ | $t_{[i,j]}^z$ | $t_{[i,j]}^{xz}$ |
|---|---|---|---|---|---|
| 1 | 0 | 0 | 1.098 (0.891) | 0.376 (0.358) | |
| | 1 | 0 | -0.480 (-0.493) | -0.100 (-0.127) | 0.226 (0.240) |
| | 1 | 1 | 0.070 (0.068) | -0.012 (-0.027) | |
| | 2 | 0 | -0.061 (-0.071) | -0.012 (-0.022) | 0.031 (0.039) |
| 2 | 0 | 0 | 1.157 | 0.277 | |
| | 1 | 0 | -0.477 | -0.087 | 0.211 |
| | 1 | 1 | 0.071 | -0.010 | |
| | 2 | 0 | -0.060 | -0.012 | 0.031 |
| Interlayer | 0 | 0 | | -0.613 (-0.678) | |
| | 1 | 0 | | | -0.030 |
| 3 | 0 | 0 | 1.148 | 0.246 | |
| | 1 | 0 | -0.478 | -0.086 | 0.209 |
| | 1 | 1 | 0.071 | -0.010 | |
| | 2 | 0 | -0.059 | -0.012 | 0.030 |
| 4 | 0 | 0 | 0.959 | 0.526 | |
| | 1 | 0 | -0.481 | -0.143 | 0.245 |
| | 1 | 1 | 0.068 | -0.018 | |
| | 2 | 0 | -0.060 | -0.012 | 0.025 |
| Interlayer | 0 | 0 | | -0.589 | |
| | 1 | 0 | | | -0.068 |
| Interbilayer | 0 | 0 | | -0.028 | |

The crystal field splitting between $d_{x^2-y^2}$ and $d_{z^2}$ orbitals ($\Delta\epsilon = \epsilon_x - \epsilon_z$) in inner layers (Layers 2 and 3; $\Delta\epsilon$ = 0.880, 0.902 eV) is larger than that in outer layers (Layers 1 and 4; $\Delta\epsilon$ = 0.722, 0.433 eV), a trend consistent with free-standing slab models [62]. However, substrate interactions introduce a notable asymmetry: the splitting in Layer 1 (0.722 eV) is significantly enhanced compared to Layer 4 (0.433 eV). The reduced crystal field splitting in Layer 4 stems primarily from strain relaxation on the top surface, which leads to a pronounced elongation of apical Ni-O bonds, particularly the $d_o^1$ bond length as listed in Table I. Except for Layer 4, these splitting magnitudes substantially exceed those reported for the bulk high-pressure *Fmmm* (0.533 eV at 21.6 GPa) and *I4/mmm* (0.547 eV at 60.3 GPa) phases [25], slightly suppressing the interorbital hybridization $t_{[10]}^{xz}$. Regarding the Ni-$d_{x^2-y^2}$ hopping parameters, compressive strain (-0.480 eV) enhances the intralayer nearest-neighbor amplitude by 3.75% relative to free-standing films (-0.462 eV), yet remains 2.71% smaller than in the high-pressure bulk (-0.493 eV). Crucially, the interlayer hopping between adjacent $d_{z^2}$ orbitals is attenuated in the bottom bilayer (Layers 1 and 2). This suppression reduces the $d_{z^2}$ bandwidth, impacting both $d_{z^2}$ bonding and antibonding states. Consequently, the distinctively flat, low-energy $\gamma$ band is derived predominantly from the interface-proximal Layers 1 and 2.

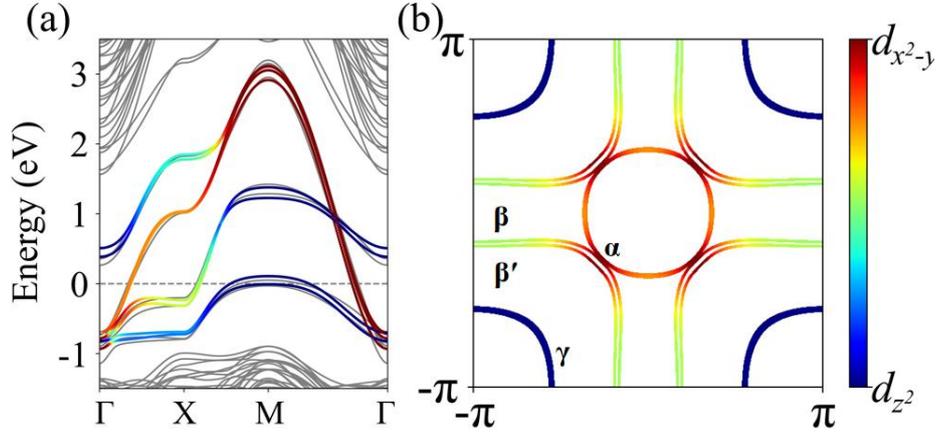

**Figure 4** Electronic structure of the two-orbital model for 1 UC LNO. (a) Orbital-projected band structure along high-symmetry directions, with Ni-$d_{x^2-y^2}$ and Ni-$d_{z^2}$ contributions shown in red and blue, respectively. Gray curves represent the original DFT+$U$ ($U$ = 3.5 eV) bands. (b) Corresponding Fermi surface. In both panels, $E_F$ is referenced to 0 eV.

Figure 4 presents the band structure and Fermi surface derived from the 1 UC two-orbital model. The Fermi surface comprises three electron pockets ($\alpha$, $\beta$, $\beta'$) exhibiting mixed orbital characters, and one hole pocket ($\gamma$) dominated by Ni-$d_{z^2}$ states. Notably, the Ni-$d_{z^2}$ bonding bands exhibit a splitting along the X-$\Gamma$ and M-$\Gamma$ directions due to the structural inequivalence induced by the substrate. We further analyze the interlayer hopping $t^z_{AB,[0,0]}$, mediated by apical O-$2p_z$ orbitals between Ni-$3d_{z^2}$ states. In Stack 1, the interlayer hopping of the $d_{z^2}$ orbital reaches -0.613 eV, which is 1.04 times larger than in Stack 2 (-0.589 eV) and 1.28 times the intralayer nearest-neighbor hopping $t^x_{[1,0]}$ (-0.480 eV), indicating interlayer hopping dominance in the bottom bilayer. Compared to the high-pressure bulk *Fmmm* phase, $t^z_{12,[0,0]}$ decreases by 9.6%, owing to the out-of-plane lattice expansion driven by in-plane substrate strain. In the strong coupling limit with large Hubbard-$U$, the interlayer spin superexchange coupling for $d_{z^2}$ orbitals $J_\perp^z$ can be estimated by $J \approx 4t^2/U$. Accordingly, the estimated $J_\perp^z$ is reduced by approximately 18% compared with that in high-pressure bulk La$_3$Ni$_2$O$_7$. Combined with the reduced Ni-$d_{z^2}$ DOS at $E_F$ in 1 UC film, the weakening of interlayer superexchange coupling is expected to suppress superconductivity. Our results provide a consistent explanation for the experimentally observed reduction in $T_c$ in compressively strained films compared to pressurized bulk samples.

## 4 Conclusions

In summary, we have performed a comprehensive, fully self-consistent first-principles investigation of LNO/SLAO(001) heterostructures across a range of film thicknesses (0.5-5 UC). Our study transcends simplified free-standing approximations by explicitly incorporating substrate-induced compressive strain and distinct interfacial reconstruction geometry. We demonstrate that Sr interdiffusion across the interface acts as an intrinsic chemical doping mechanism, which is essential for maintaining charge neutrality and stabilizing the experimentally observed hole doping. Our DFT+$U$ calculations reveal an electronic topology strikingly similar to that of the superconducting high-pressure bulk phase. Crucially, while typical full-UC models successfully capture the $\alpha$ and $\beta$ electron sheets, the 1 UC configuration also notably reproduces the $\gamma$ hole pocket at the Fermi surface, thereby achieving the highest fidelity to ARPES measurements. Furthermore, layer-resolved analysis identifies that the specific $\gamma$ band residing closest to $E_F$ originates exclusively from the interface-proximal Stack 1. This implies that the macroscopic conductivity is dominated by the interfacial bilayer, a finding consistent with experimental evidence attributing the robust superconductivity to the high quality of the interface layer. Motivated by the accurate reproduction of the experimental Fermi surface topology, we finally constructed a minimal double-stacked two-orbital tight-binding model based on the representative 1 UC system to capture the essential low-energy physics. Our analysis suggests that the reduction in $T_c$ in thin films relative to the bulk stems from the synergistic suppression of two critical factors: the lower Ni-$d_{z^2}$ DOS at $E_F$ and the weakened out-of-plane superexchange interaction ($J_\perp^z$) due to reduced interlayer hopping. Collectively, this work establishes interface engineering as a decisive factor in modulating the orbital physics of nickelates, providing a rigorous theoretical framework for understanding the critical role of the $\gamma$ band and guiding the design of future high-$T_c$ thin-film devices.

## Acknowledgments

This work was supported by National Key R&D Program of China (Nos. 2022YFA1402304, 2022YFA1402802, and 2023YFA1406200), National Natural Science Foundation of China (Grants Nos. 12494591, 12274169, 92165204, and 92565303), Guangdong Provincial Key Laboratory of Magnetoelectric Physics and Devices (Grant No. 2022B1212010008), Research Center for Magnetoelectric Physics of Guangdong Province (2024B0303390001), Guangdong Provincial Quantum Science Strategic Initiative (GDZX2401010), and the Fundamental Research Funds for the Central Universities. Some of the calculations were performed at the High Performance Computing Center of Jilin University and using TianHe-1(A) at the National Supercomputer Center in Tianjin.